\documentclass[12pt]{article}
\input epsf
\newcommand{\be}{\begin{eqnarray}}
\newcommand{\ee}{\end{eqnarray}}
\newcommand{\Qt}{\tilde{Q}}
\newcommand{\z}{\zeta}
\newcommand{\abs}[1]{\left| #1 \right|}
\newcommand{\half}{\frac{1}{2}}
\newcommand{\C}{{\sf C\hspace*{-0.9ex}%
    \rule{0.15ex}{1.3ex}\hspace*{0.9ex}}}
\begin{document}

\begin{titlepage}

\begin{flushright}
hep-th/9901091
\end{flushright}

\begin{center}
\vskip3em
{\large\bf Branes at Generalized Conifolds and Toric Geometry}

\vskip3em
{\ R.\ von Unge\footnote{E-mail:unge@physics.muni.cz}}\\ \vskip .5em
{\it Department of Theoretical Physics and Astrophysics\\
Faculty of Science, Masaryk University\\
Kotl\'{a}\v{r}sk\'{a} 2, 611 37, Brno\\
Czech Republic}

\vskip2em
\end{center}

\vfill

\begin{abstract}
We use toric geometry to investigate the recently proposed relation
between a set of D3 branes at a generalized conifold singularity and
type IIA configurations of D4 branes stretched between a number of
relatively rotated NS5 branes. In particular we investigate how various
resolutions of the singularity corresponds to moving the NS branes and
how Seiberg's duality is realized when two relatively rotated
NS-branes are interchanged.
\end{abstract}

\vfill
\end{titlepage}

\section{Introduction}
The duality between ${\cal N}=4$ supersymmetric Yang-Mills theory and
supergravity on the space $AdS_{5}\times S^{5}$ recently conjectured
by Maldacena \cite{malda} was found by studying the physics and the
near horizon geometry of D3-branes in flat space. Therefore, to try to
extend this conjecture it is natural to study D3-branes on more
complicated spaces. The most natural generalization is perhaps to
study D3-branes at an orbifold singularity \cite{n2orb,orbres}.

In this context, it is also useful to study T-dual versions of these
models. Under T-duality the singularity gets mapped into a
configuration of NS5-branes and the D3-branes gets mapped into
D4-branes. The so called Brane Box \cite{bb1,bb2} models are related
by two T-duality transformations to D3-branes at an orbifold
singularity $\C^3 /\Gamma$. The T-duality transformations maps parameters
and moduli of the field theory of the branes into geometrical
quantities of the T-dual brane configuration which makes certain
phenomena more easily studied. One can for instance realize Seiberg's
duality as a ``reshuffling'' of branes \cite{sd} and in some cases
solve the field theory by lifting the brane configuration to M-theory
\cite{witsol}.

More recently, following the work by Klebanov and Witten
\cite{igored}, D3-branes on non-orbifold singularities, leading to
gauge theories which are not just projections of ${\cal N}=4$ theory,
have been studied
\cite{extra1,sphor,extra2,glr,dm,bost,elopez,extra3,uran}. The basic
example of \cite{igored} concerned D3-branes on a conifold singularity
and the subsequent work is various generalizations of that.

One interesting article is \cite{uran} where D3-branes on a quotient
of the conifold by an appropriate discrete isometry group were
studied. A duality between D3-branes on these singular spaces and
configurations of NS-branes and D4-branes in type IIA theory was also
proposed and the relation between resolutions of the singularity and
movement of the branes was studied. Furthermore it was proposed that
Seiberg's duality (the interchanging of NS and NS$'$ branes in this
context) could be realized as flop transitions between topologically
different small resolutions of the singularities.

The purpose of this paper is to use toric geometry (for an
introduction see \cite{agm}) to study this correspondence in more
detail and thereby provide additional evidence for the proposed
duality and the correspondence between the resolution of the
singularity and the movement of the branes. Using toric methods we
will be able to show that the Higgs branch of the moduli space of the
gauge theory is indeed the generalized conifold singularity. We will
also be able to get an explicit handle on the various resolutions of
this singularity by including the FI-terms in the calculation.

Toric geometry has been used previously in the study of 3-branes at
various singularities. The methods used in this paper were introduced
in \cite{orbres} where D3-branes on $\C^3 /Z_{k}$ orbifolds were
studied. In particular it was shown that only geometric phases of the
singularity was seen by the D3-branes. More complicated orbifold
singularities were studied in \cite{topch2,topch,cfdb,sphor},
including conifiold singularities as subsingularities. It was also shown
that the D-branes are sensitive to topologically different resolutions
of the singularities, differing by flop transitions.

This paper is organized as follows: section \ref{sec:basic} studies
the simplest non-trivial example and explicitly derives the toric data
describing this singularity. In section \ref{sec:resls} we use the
toric data to study how various resolutions are related to movement of
the NS-branes in the T-dual version of the model. In section
\ref{sec:general} we study a more complicated example corresponding to
the space $xy=z^3 w^3 $ and use that to make claims about the general
correspondence between resolutions of the singularity and movement of
the NS-branes. In section \ref{sec:seib} we use our methods to study
a situation related to Seiberg's duality as was proposed in
\cite{uran} and in section \ref{sec:concl} we summarize the paper and
discuss some directions for future research. The paper is finished
with an appendix which contains the toric data for some of the more
complicated examples studied in the paper.

\section{The basic example}
\label{sec:basic}
We are interested in studying $N$ D3-branes on a singularity of the type
$xy=z^{k}w^{k'}$. According to \cite{uran} this is T-dual to a
configuration with $N$ D4-branes wound around a compact direction
($x_6$ in our case) and stretched between $k$ NS-branes and $k'$
NS$'$-branes which are placed at various points along the circle. The
branes are oriented as follows
\be
\begin{array}{lcccccccccc}
& 0 & 1 & 2 & 3 & 4 & 5 & 6 & 7 & 8 & 9 \\
NS & \times & \times & \times & \times & \times & \times & - & - & - & - \\
NS' & \times & \times & \times & \times & - & - & - & - & \times & \times \\
D4 & \times & \times & \times & \times & - & - & \times & - & - & - 
\end{array},
\ee
where a cross means of infinite extent and a dash means point like in
the particular dimension.

\begin{figure}[thb]
 \begin{center}
 \mbox{\epsfysize=5cm\epsfbox{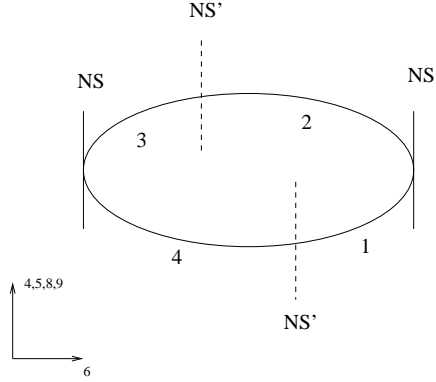}} %specifies the size and filename
  \caption{The brane configuration}\label{fig:NSconf}
  \end{center}
\end{figure}

Let us study the simplest non-trivial example, namely the
configuration with two NS-branes and two NS$'$-branes given in figure
\ref{fig:NSconf} and, according to \cite{uran}, T-dual to D3-branes on
a singularity of the type $xy=z^{2}w^{2}$.
The D-term equations for this model are 
\be\label{Dterm}
\abs{Q_{41}}^{2} - \abs{\Qt_{14}}^{2} - \abs{Q_{12}}^{2} +
  \abs{\Qt_{21}}^{2} &=& \z_1, \nonumber\\
\abs{Q_{12}}^{2} - \abs{\Qt_{21}}^{2} - \abs{Q_{23}}^{2} +
  \abs{\Qt_{32}}^{2} &=& \z_2, \nonumber\\
\abs{Q_{23}}^{2} - \abs{\Qt_{32}}^{2} - \abs{Q_{34}}^{2} +
  \abs{\Qt_{43}}^{2} &=& \z_3, \\
\abs{Q_{34}}^{2} - \abs{\Qt_{43}}^{2} - \abs{Q_{41}}^{2} +
  \abs{\Qt_{14}}^{2} &=& \z_4, \nonumber
\ee
and the superpotential is
\be
W \propto \Qt_{21}Q_{12}Q_{23}\Qt_{32} -
          \Qt_{32}Q_{23}Q_{34}\Qt_{43} +
          \Qt_{43}Q_{34}Q_{41}\Qt_{14} -
          \Qt_{14}Q_{41}Q_{12}\Qt_{21},
\ee
giving the F-term constraints
\be
Q_{12}\Qt_{21} &=& Q_{34}\Qt_{43}, \nonumber\\
Q_{23}\Qt_{32} &=& Q_{41}\Qt_{14}.
\ee
We can solve the F-term constraints in terms of a minimal set of
fields. If we choose to have $\Qt_{43}$ and $\Qt_{14}$ as dependent on
the other fields, the solution can be represented as follows
\be \label{Mmatrix}
\begin{array}{lcccccc}
        & Q_{12} & \Qt_{21} & Q_{23} & \Qt_{32} & Q_{34} & Q_{41}\\
Q_{12}  & 1      & 0        & 0      &  0       & 0      & 0\\
\Qt_{21}& 0      & 1        & 0      &  0       & 0      & 0\\
Q_{23}  & 0      & 0        & 1      &  0       & 0      & 0\\
\Qt_{32}& 0      & 0        & 0      &  1       & 0      & 0\\
Q_{34}  & 0      & 0        & 0      &  0       & 1      & 0\\
\Qt_{43}& 1      & 1        & 0      &  0       & -1     & 0\\
Q_{41}  & 0      & 0        & 0      &  0       & 0      & 1\\
\Qt_{14}& 0      & 0        & 1      &  1       & 0      &-1
\end{array}.
\ee
To be able to put the F-term constraints and the D-term constraints on
equal footing (as a symplectic quotient) we will introduce homogeneous
coordinates $p_0 \ldots p_7$ as follows
\be
\begin{array}{cccc}
Q_{12} = p_3 p_4 & Q_{23} = p_2 p_6 & Q_{34} = p_3 p_5 & Q_{41} = p_2 p_7\\
\Qt_{21}=p_1 p_5 & \Qt_{32}=p_0 p_7 & \Qt_{43}=p_1 p_4 & \Qt_{14}=p_0 p_6
\end{array},
\ee
in which the F-term constraints are automatically satisfied. The
homogeneous coordinates span a $\C^8$ which is acted upon by a $U(1)^2$
action under which the original $Q$ and $\Qt$ fields are the invariant
coordinates. The solution can also be represented in the form of the
matrix 
\be \label{Tmatrix}
T = \left(\begin{array}{cccccccc}
p_0 & p_1 & p_2 & p_3 & p_4 & p_5 & p_6 & p_7 \\
0 & 0 & 0 & 1 & 1 & 0 & 0 & 0 \\
0 & 1 & 0 & 0 & 0 & 1 & 0 & 0 \\
0 & 0 & 1 & 0 & 0 & 0 & 1 & 0 \\
1 & 0 & 0 & 0 & 0 & 0 & 0 & 1 \\
0 & 0 & 0 & 1 & 0 & 1 & 0 & 0 \\
0 & 0 & 1 & 0 & 0 & 0 & 0 & 1
\end{array}\right),
\ee
where to find which power of $p_i$ a particular $Q$ contains, one
takes the corresponding column in (\ref{Tmatrix}) and take the scalar
product with the corresponding row in (\ref{Mmatrix}).

\begin{figure}[htb]
 \begin{center}
 \mbox{\epsfysize=5cm\epsfbox{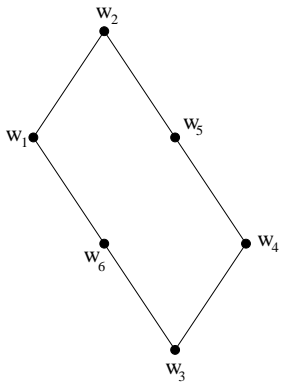}} %specifies the size and filename
  \caption{The singularity}\label{fig:sing}
  \end{center}
 \end{figure}

The F-term constraints can now be represented as a symplectic quotient
on the space spanned by the homogeneous coordinates. The symplectic
quotient is implemented by the previously mentioned $U(1)^{2}$ action
under which the homogeneous coordinates has the following charges
\be
\left(\begin{array}{cccccccc}
p_0 & p_1 & p_2 & p_3 & p_4 & p_5 & p_6 & p_7 \\
0 & -1 & 0 & -1 & 1 & 1 & 0 & 0 \\
-1 & 0 & -1 & 0 & 0 & 0 & 1 & 1
\end{array}\right).
\ee
Now we have to find how the ordinary D-term constraints will act on
the homogeneous coordinates. To do this we introduce the matrix $U$
defined by $T U^{tr} = Id$
\be
U = \left(\begin{array}{cccccccc}
0 & 0 & 0 & 0 & 1 & 0 & 0 & 0 \\
0 & 1 & 0 & 0 & 0 & 0 & 0 & 0 \\
0 & 0 & 0 & 0 & 0 & 0 & 1 & 0 \\
1 & 0 & 0 & 0 & 0 & 0 & 0 & 0 \\
0 & 0 & 0 & 1 & -1 & 0 & 0 & 0 \\
0 & 0 & 1 & 0 & 0 & 0 & -1 & 0
\end{array}\right).
\ee
Then the D-term constraints are represented by the charge matrix $VU$
where $V$ contains the charges of the basic fields under the
particular gauge group. In our case we have three independent gauge
groups (the charges of the fourth one is given in terms of the other
three) so $V$ is given by
\be
V = \left(\begin{array}{cccccc}
Q_{12} & \Qt_{21} & Q_{23} & \Qt_{32} & Q_{34} & Q_{41} \\
-1 & 1 & 0 & 0 & 0 & 1 \\
1 & -1 & -1 & 1 & 0 & 0 \\
0 & 0 & 1 & -1 & -1 & 0
\end{array}\right),
\ee
giving us a charge matrix
\be
VU = \left(\begin{array}{cccccccc}
0 & 1 & 1 & 0 & -1 & 0 & -1 & 0 \\
1 & -1 & 0 & 0 & 1 & 0 & -1 & 0 \\
-1 & 0 & 0 & -1 & 1 & 0 & 1 & 0
\end{array}\right).
\ee
Now we concatenate the charge matrix from the F-term constraints with
the charge matrix representing the D-term constraints to get the full
reduction on the $\C^8$ spanned by the homogeneous coordinates. The
cokernel of the transpose of this matrix gives us the toric data for
the space of interest \cite{orbres}. In our case it is given by
\be
\tilde{T} = \left(\begin{array}{cccccccc}
 1 &  0 &  1 & 0 & 0 & 0 & 1 & 1 \\
-1 & -1 &  1 & 1 & 0 & 0 & 0 & 0 \\
 1 &  2 & -1 & 0 & 1 & 1 & 0 & 0
\end{array}\right).
\ee
All of these vectors lie with their tip on the plane with normal
$(1,1,1)$ at a distance of $1/\sqrt{3}$ from the origin. With the
notation
\be
w_1 = (1,-1,1) & w_2 = (0,-1,2) & w_3 = (1,1,-1) \nonumber\\
w_4 = (0,1,0)  & w_5 = (0,0,1)  & w_6 = (1,0,0)
\ee
we can draw a picture (figure \ref{fig:sing}) of where the vectors hit
the plane
in which we recognize the toric description of the space $xy=z^2 w^2 $
which is in agreement with the proposal of \cite{uran}.

As usual, when we include the FI-terms we resolve the space. Because
we have no FI-parameters for the F-term constraints not all possible
phases are realized and only geometrical phases are seen exactly as in
\cite{orbres}. To show this and to further study these resolutions we
therefore include the FI-parameters and give the charge matrix in a
particularly useful form.
\be
\left(\begin{array}{ccccccccc}
0 & 0 & 0 & 0 & 0 & 0 & -1 & 1 & \z_1 + \z_2 \\
0 & 0 & 0 & 0 & 1 & -1 & 0 & 0 & \z_2 + \z_3 \\
0 & 1 & 1 & 0 & -1 & 0 & -1 & 0 & \z_1 \\
1 & -1 & 0 & 0 & 1 & 0 & -1 & 0 & \z_2 \\
-1 & 0 & 0 & -1 & 1 & 0 & 1 & 0 & \z_3
\end{array}\right).
\ee
The first two rows can be used to eliminate some of the homogeneous
coordinates for different values of $\z_1 + \z_2$ and $\z_2 +
\z_3$. The result is that only the geometric phases are realized.

For reference we give the data for the different type of singularities
here. These can all be obtained from the matrix above by invertible
row operations and we have assumed that $\z_1 + \z_2 \geq 0$ and $\z_2
+ \z_3 \geq 0$. We also give the pictures showing which singularity is
controlled by which parameter. The solid lines indicate the
singularity and the dotted lines represent the resolved part.

The data for the conifold singularities (of the type $xy=zw$) is 
\be\label{conmat}
\left(\begin{array}{ccccccccc}
0 & 0 & 1 & -1 & 0 & 1 & -1 & 0 & \z_1 \\
1 & 0 & 0 & 1 & 0 & -1 & -1 & 0 & \z_2 \\
-1 & 1 & 0 & 0 & 0 & -1 & 1 & 0 & \z_3 \\
0 & -1 & -1 & 0 & 0 & 1 & 1 & 0 & \z_4
\end{array}\right),
\ee
corresponding to the pictures
  \begin{center}
 \mbox{\epsfxsize=11cm\epsfbox{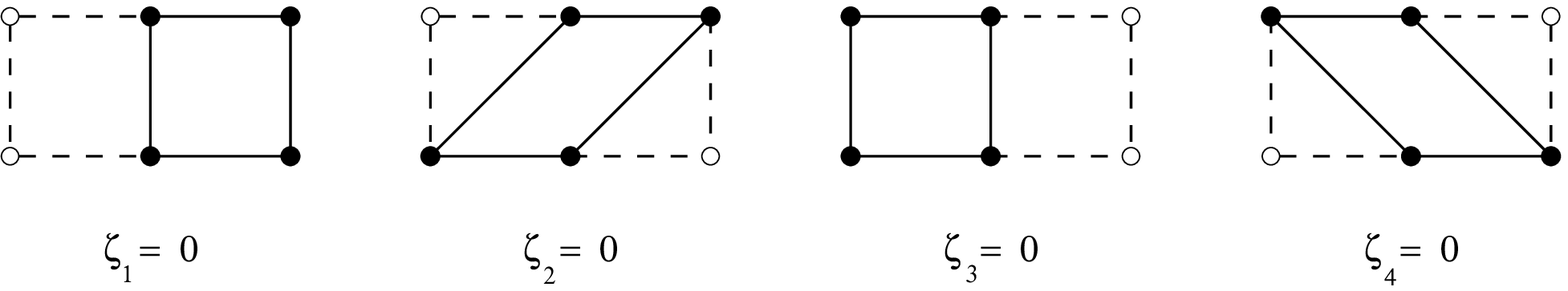}} %specifies the size and filename
  \end{center}
for the orbifold singularities (of the type $xy=z^2$) it is
\be
\left(\begin{array}{ccccccccc}
1 & 0 & 1 & 0 & 0 & 0 & -2 & 0 & \z_1 + \z_2\\
0 & 1 & 0 & 1 & 0 & -2 & 0 & 0 & \z_2 + \z_3
\end{array}\right),
\ee
corresponding to the pictures
  \begin{center}
 \mbox{\epsfysize=2cm\epsfbox{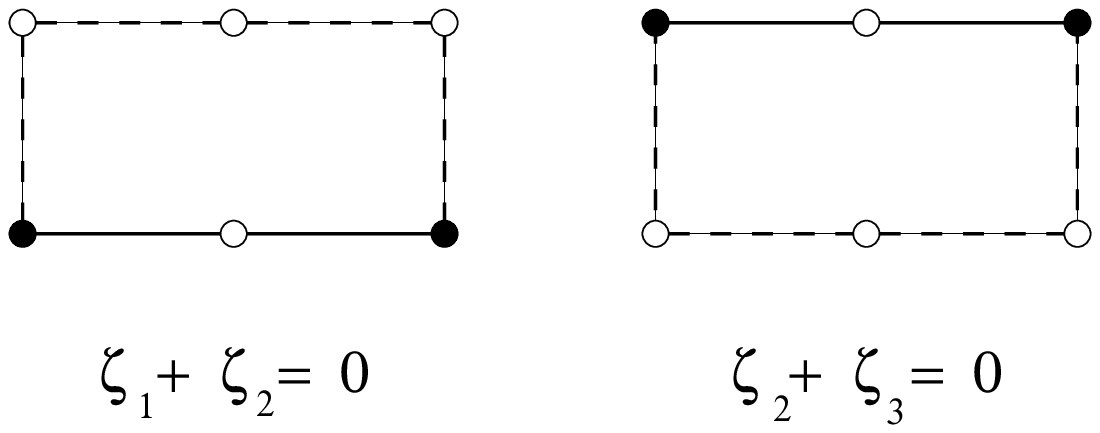}} %specifies the size and filename
  \end{center}
and finally for the suspended pinch point singularities (of the type
$xy=zw^{2}$) it is
\be
\left(\begin{array}{ccccccccc}
-1 & 0 & 1 & -2 & 0 & 2 & 0 & 0 & \z_1 - \z_2 \\
2 & -1 & 0 & 1 & 0 & 0 & -2 & 0 & \z_2 - \z_3 \\
-1 & 2 & 1 & 0 & 0 & -2 & 0 & 0 & \z_3 - \z_4 \\
0 & -1 & -2 & 1 & 0 & 0 & 2 & 0 & \z_4 - \z_1 
\end{array}\right),
\ee
corresponding to the pictures
  \begin{center}
 \mbox{\epsfxsize=11cm\epsfbox{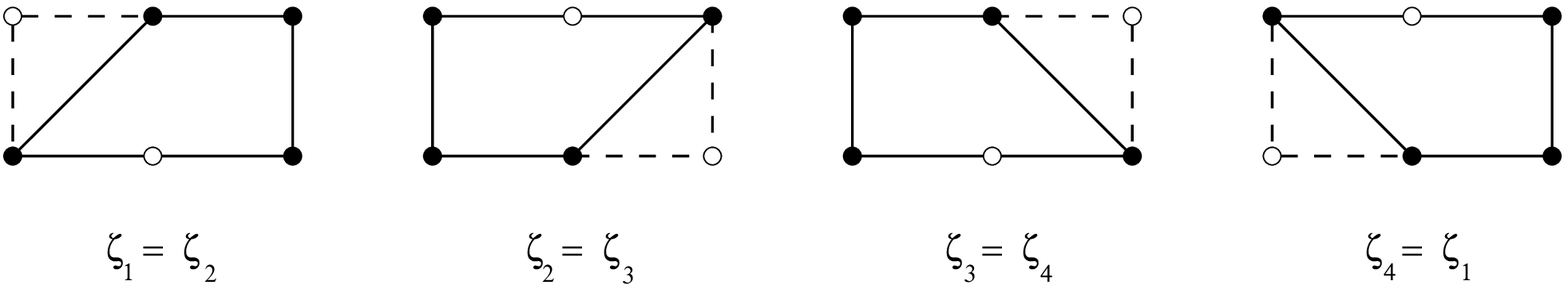}} %specifies the size and filename
  \end{center}

\section{Analysis of subsingularities}
\label{sec:resls}
Using these expressions we see that for particular values of the
FI-parameters we will have various unresolved subsingularities in our
configuration. It is interesting to analyze how the brane picture
corresponds to the singularity picture for these cases.

If we start with the orbifold singularities we see that if we for
example take $\z_1+\z_2 = 0$ but leave $\z_3$ arbitrary we will leave
the first one of them unresolved. Looking at the D-term equations and
assuming $\z_1$ and $\z_3$ are positive we see that they are satisfied
if we give expectation values to the chiral fields as follows
\be
 \Qt_{21} &=& \z_1^{\half}, \nonumber\\
 \Qt_{43} &=& \z_3^{\half}.
\ee
These expectation values corresponds to moving both the NS-branes in
the $x_7$-direction. This will break the first and the second gauge
group down to their diagonal component and similarly for the third and
the fourth gauge group. Furthermore, if we insert these expectation
values in the superpotential we see that the fields $Q_{12}$ and
$Q_{34}$ becomes the adjoint fields of the unbroken gauge groups and
we get the usual elliptic $N=2$ model with two gauge groups
\cite{witsol}.

A slightly more complicated example is to study what happens when we
leave one of the conifold singularities unresolved. By studying the
matrix (\ref{conmat}) we see that this happens when we for instance
put $\z_1 = 0$. Inspection of the D-term constraints reveals that they
are satisfied if we put
\be
 \Qt_{32} &=& \z_2^{\half},\nonumber\\
 \Qt_{43} &=& \left( \z_2 + \z_3 \right)^{\half},
\ee
which corresponds to moving the second NS-brane and the second
NS$'$-brane in the $x_7$-direction. When we insert this into the
superpotential we get
\be
\lefteqn{W \propto \z_{2}^{\half} \Qt_{21}Q_{12}Q_{23} -
\z_{2}^{\half}\left(\z_2 + \z_3 \right)^{\half} Q_{23} Q_{34} +}&
\\
& \left( \z_2 + \z_3 \right)^{\half}
 Q_{34}Q_{41}\Qt_{14} - \Qt_{14}Q_{41}Q_{12}\Qt_{21},
\nonumber
\ee
This will give a mass to the fields $Q_{23}$ and $Q_{34}$,
and integrating them out we find
\be
W \propto
 Q_{41}\Qt_{14}\Qt_{21}Q_{12} - \Qt_{14}Q_{41}Q_{12}\Qt_{21},
\ee
which is indeed the superpotential for the conifold as proposed in
\cite{igored}.

We can also ask what will happen if we leave one of the suspended
pinch point singularities unresolved.  We see that this will happen if
we choose $\z_1 =\z_2$. However, we are still free to choose $\z_1 +
\z_2 $. Let us study the simplest possibility $\z_1 +\z_2 =0$
first. This means that $\z_1 = \z_2 =0$ and we can assume that $\z_3 >
0$. By inspection we find that the D-term constraints are satisfied if
we choose
\be
 \Qt_{43} = \z_{3}^{\half},
\ee
which corresponds to moving the second NS-brane in the
$x_7$-direction. Again we break the third and fourth gauge groups to the
diagonal combination and $Q_{34}$ becomes a chiral superfield in the
adjoint of the unbroken part of that group. The superpotential becomes
\be
W \propto
  \Qt_{21}Q_{12}Q_{23}\Qt_{32} - \Qt_{14}Q_{41}Q_{12}\Qt_{21}
  +\z_{3}^{\half}\left( \Qt_{14}Q_{34}Q_{41} - Q_{23} Q_{34}
  \Qt_{32}\right),
\ee
which is the superpotential one would get from a model
with two NS$'$-branes and only one NS-brane.

If we in the last example choose $\z_1 +\z_2 > 0$ instead we still
keep the suspended pinch point singularity unresolved. However, we
will resolve some of its subsingularities giving us a more complicated
situation. The D-term equations tells us that we have to give
expectation values to three of the hypermultiplets which seems to be a
highly nontrivial situation. How come the suspended pinch point
singularity is still unresolved?  The answer can be found if one uses
that we know that the FI-parameter for a particular gauge group is
geometrically encoded as the difference in the $x_7$ coordinate of the
NS-brane to the right of of the gauge group and to the left of the
gauge group \cite{sd}. From this follows that the final
configuration still has two NS$'$-branes and one NS-brane in a line,
although this line is slightly tilted in the $x_6 ,x_7$ plane.

\bigskip
We may also reverse the question and ask what happens with the
singularity when we move a particular NS-brane or NS$'$-brane in a
particular way instead of asking what happens with the brane
configuration when we resolve the singularity.  Using the
correspondence between FI-parameters and the relative $x_7$
coordinates of the branes again, it is possible to find out exactly
what FI-parameters a particular movement corresponds to. For instance,
if we move the NS-brane separating the first and the second gauge
groups a distance $\z$ in the negative $x_7$ direction, we will get
the following FI-parameters
\be
 \z_1 &=& -\z, \nonumber\\
 \z_2 &=& \z, \\
 \z_3 &=& 0, \nonumber\\
 \z_4 &=& 0, \nonumber
\ee
which, using the D-term equations, corresponds to giving an expectation
value to $Q_{12}=\z^{\half}$.
Since $\z_1 + \z_2 \geq 0$ and $\z_2 + \z_3 \geq 0$ we can use the
previous formulas to see that the unresolved singularities are the
third suspended pinch point singularity and its subsingularities.

\section{More examples}
\label{sec:general}
\begin{figure}[htb]
 \begin{center}
 \mbox{\epsfysize=5cm\epsfbox{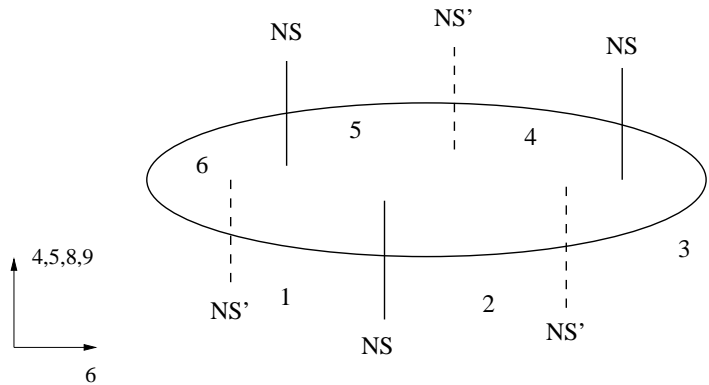}} %specifies the size and filename
  \caption{A more complicated example}\label{fig:3NS3NSp}
  \end{center}
 \end{figure}

After studying the simplest example in detail we may now turn to
investigate more general models such as the one with 3 NS-branes and 3
NS$'$-branes as shown in figure \ref{fig:3NS3NSp}.
We can use the method described above to solve for the Higgs branch of
the moduli space of the field theory describing this brane
configuration. The toric data and the charge matrix giving this moduli
space as a symplectic quotient is given in formula (\ref{Qfor3NS}) in
the appendix. The toric data corresponds to a singularity of the type
$xy=z^{3}w^{3}$ as expected from the arguments given in
\cite{uran}. We can use the charge matrix to investigate what phases
are realized when we put 3-branes on such a singularity and also to
find out how moving the NS- and NS$'$-branes are related to various
resolutions of the singularity. In particular, there are four
different ways of moving the NS- and NS$'$-branes in the $x_7$
direction (turning on FI-terms). Namely, we can move an NS-brane in
the negative $x_7$ direction, we can move an NS$'$-brane in the
negative $x_7$ direction, we can move an NS-brane in the positive
$x_7$ direction and finally we can move an NS$'$-brane in the positive
$x_7$ direction. By using the charge matrix (\ref{Qfor3NS}) given in
the appendix we see that these four distinct ways of moving the branes
correspond to four distinct ways of resolving the singularity. These
are given in the figure below in the order corresponding to the
movements given in the text.
  \begin{center}
 \mbox{\epsfxsize=12cm\epsfbox{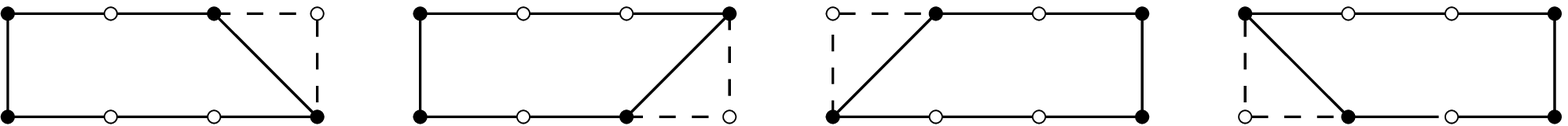}} %specifies the size and filename
  \end{center}
It should be pointed out that it does not matter which NS-brane or
NS$'$-brane one is moving for this correspondence to hold. The only
thing that matters is if it is an NS-brane or an NS$'$-brane or if we
move it in the positive or negative $x_7$ direction. We may also check
that this is true for the previously studied configuration given in
figure \ref{fig:NSconf}.

\section{Seiberg's duality}
\label{sec:seib}
Finally we would like to study what happens to the singularity when
one moves an NS-brane past an NS$'$-brane in the $x_6$ direction. This
should be related to Seiberg's duality as was observed in
\cite{sd}. In \cite{uran} this fact was used to conjecture that
Seiberg's duality in the ``D4-branes between NS-branes'' picture is related
to flop transitions between topologically distinct small resolutions of the
singularity in the T-dual picture. We will now investigate this claim
using toric methods.

We can study this phenomenon (following \cite{sd})
without encountering the singularity that would result from actually
letting the NS-brane and the NS$'$-brane meet in space by letting one
of the branes move in the $x_7$ direction before moving it in the
$x_6$ direction. Then we are free to move the brane in the $x_6$
direction until we pass the NS$'$-brane since they are at different
points in $x_7$ and after the branes have passed each other we can let
the branes move back to their original positions in $x_7$ giving us a
configuration where the NS and NS$'$-brane has changed places.

\begin{figure}[htb]
 \begin{center}
 \mbox{\epsfysize=5cm\epsfbox{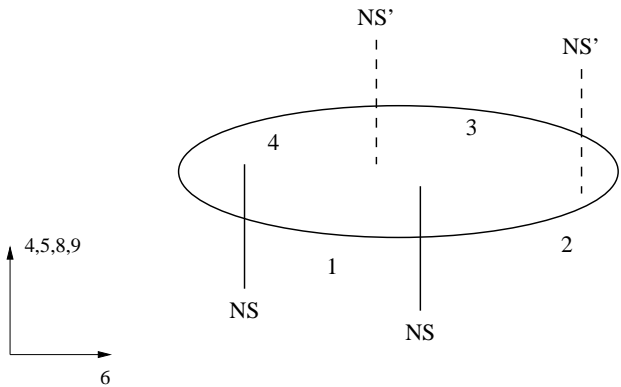}} %specifies the size and filename
  \caption{The dual configuration}\label{fig:dual}
  \end{center}
 \end{figure}

{}From the previous discussion we know what happens when we
move a brane in the $x_7$ direction and the singularity structure does
not depend on the $x_6$ coordinates of the NS-branes so what we have
to study is what happens when we move the NS-brane in the $x_7$
coordinate in the configuration where one NS-brane and one NS$'$-brane
has changed places. To be concrete, let us therefore investigate the
configuration with two NS-branes and two NS$'$-branes related to the
previously studied configuration by an interchange of the NS-brane and
the NS$'$-brane surrounding the fourth gauge group. This configuration
is given in figure \ref{fig:dual}.
The field theory corresponding to this configuration also has Coulomb
branches (corresponding to moving the D4-branes between the NS-branes
or between the NS$'$-branes) but since we are only interested in the
Higgs branches we will set the vacuum expectation value of the adjoint
chiral superfield which exists for the first and third gauge groups to
zero. Doing this and repeating the previous analysis we get the same
toric data as in the first example. However, the charge matrix is
slightly different and is given in the appendix. In particular, the
parameters that control the sizes of the orbifold singularities are
 $\z_1$ and $\z_3$.

If we repeat the analysis performed in section \ref{sec:resls} for the
configuration in figure \ref{fig:NSconf} we find that the same general
results hold. The singularity is resolved in the same way, for
instance, if we move any one of the NS-branes in the positive $x_7$
direction we will remove an upper left triangle from the toric diagram
just as for the original configuration. However, the FI-parameters
that control the sizes of the various singularities are now different
and we may ask if it is possible to map them into each other. To make
the question more precise let us study the original configuration
where we move the NS-brane on the left of the fourth gauge group a
distance $\z$ in the positive $x_7$ direction and the NS$'$-brane on
the right a distance $\epsilon$ in the negative $x_7$ direction. This
corresponds to introducing FI-parameters with the following values
\be
\z_1 &=& \epsilon, \nonumber\\
\z_2 &=& 0, \nonumber\\
\z_3 &=& \z, \\
\z_4 &=& -\z -\epsilon ,\nonumber
\ee
and we see that in particular $\z_1 +\z_2 \geq 0$ and $\z_2 +\z_3 \geq
0$ so we can use the formulas given in the text. On the other hand, in
the dual configuration (where this NS-brane and NS$'$-brane have
changed places) this would correspond to a set of FI-parameters
\be
\tilde{\z}_1 &=& -\z ,\nonumber\\
\tilde{\z}_2 &=& 0 ,\nonumber\\
\tilde{\z}_3 &=& -\epsilon ,\\
\tilde{\z}_4 &=& \z +\epsilon ,\nonumber
\ee
where we see that $\z_1 \leq 0$ and $\z_3 \leq 0$ which determines
which of the homogeneous coordinates to use for the dual
configuration.

Now we can compare parameters. For each of the subsingularities there
is a particular combination of FI-parameters which controls its
size. If we assume that nothing happens with the sizes when we move in
$x_6$ we can equate them and we get this system of equations (one
equation for each subsingularity)
\be
\z_1 +\z_2 &=& -\tilde{\z}_3 ,\nonumber\\
\z_2 +\z_3 &=& -\tilde{\z}_1 ,\nonumber\\
\z_1 &=& \tilde{\z}_1 +\tilde{\z}_4 ,\nonumber\\
\z_2 &=& \tilde{\z}_2 ,\nonumber\\
\z_3 &=& -\tilde{\z}_1 -\tilde{\z}_2 ,\nonumber\\
\z_4 &=& -\tilde{\z}_4 ,\\
\z_1-\z_2 &=& -2\tilde{\z}_2 -\tilde{\z}_3 ,\nonumber\\
\z_2 -\z_3 &=& \tilde{\z}_1 + 2\tilde{\z}_2 ,\nonumber\\
\z_3 -\z_4 &=& \tilde{\z}_3 +2\tilde{\z}_4 ,\nonumber\\
\z_4 -\z_1 &=& -\tilde{\z}_1 -2\tilde{\z}_4 ,\nonumber
\ee
where $\z_{i}$ are the FI-parameter in the original model and
$\tilde{\z}_{i}$ are the FI-parameter in the dual configuration. This
has a solution which is
\be \label{newsol}
\tilde{\z}_1 &=& \z_1 + \z_4 ,\nonumber\\
\tilde{\z}_2 &=& \z_2 ,\nonumber\\
\tilde{\z}_3 &=& -\z_1 -\z_2 ,\\
\tilde{\z}_4 &=& -\z_4 ,\nonumber
\ee
which means that it seems to be possible to move the NS-branes in the
$x_6$ direction while keeping the sizes of the various singularities
fixed. Furthermore, the FI-parameters in (\ref{newsol}) precisely
corresponds to the FI-parameters one would expect if one just
interchanged the NS-brane and the NS$'$-brane surrounding the fourth
gauge group. To see this we can assume that the $x_7$ positions of the
NS-branes are (with the convention that the NS$'$-brane between the
fourth and the first gauge group is the first brane)
\be
x_7 (1) &=& 0 ,\nonumber\\
x_7 (2) &=& \z_1 ,\nonumber\\
x_7 (3) &=& \z_2 + \z_1 ,\\
x_7 (4) &=& -\z_4 .\nonumber
\ee
If we interchange the the first and the fourth NS-branes the positions
would instead be
\be
x_7 (\tilde{1}) &=& -\z_4 ,\nonumber\\
x_7 (\tilde{2}) &=& \z_1 ,\nonumber\\
x_7 (\tilde{3}) &=& \z_2 + \z_1 ,\\
x_7 (\tilde{4}) &=& 0 ,\nonumber
\ee
which gives exactly the FI-parameters in (\ref{newsol}).

We thus see that there is a one to one map between the FI-parameters
between the two configuration related by interchanging of one NS-brane
and one NS$'$-brane. In particular, the various subsingularities of
the generalized conifold will have the same sizes in both
configurations. This means that in this formalism we do not see any
sign that the two configuration are related by a flop transition
between topologically different small resolutions of the singularities
as was conjectured in \cite{uran}\footnote{The real situation is
slightly more complicated however. The $x_6$ position of the branes
are encoded in the integrals over the B-field over the two-cycles of
the singularity. When we change the relative $x_6$ position of the
branes we change the B-field on the corresponding cycle and what
happens when two branes cross is that the B-field on the cycle in
question actually flips sign. What we have shown is that the real
size, as measured by the imaginary part of the complexified K\"{a}hler
form $B+iJ$, does not change when the NS-branes cross. However, since
the real part of the complexified K\"{a}hler form changes, there might
still be something like a flop transition taking place in the
complexified K\"{a}hler moduli space. It would be interesting to
continue this work along this line of reasoning. I would like to thank
A. Uranga for discussions on this topic.}.

\section{Conclusions}
\label{sec:concl}
We have shown how the toric description of the moduli space of
D3-branes at a generalized conifold singularity is related to the
T-dual version with D4-branes suspended between NS-branes and
NS$'$-branes. By studying various subsingularities of the space it was
possible to investigate how moving the NS or NS$'$ branes in the $x_7$
direction in the dual configuration resolves the singularity in
various ways.

We also showed that with toric methods we were able to study what
happens as move two relatively rotated NS-branes across each other in
the $x_6$-direction, corresponding to performing Seiberg's duality in
one of the gauge groups of the dual configuration. In \cite{uran} it
was conjectured that this should be related to flop transitions in the
toric diagram. We showed that there is a one to one map between the
FI-parameters of the original model and the Seiberg dual model and
that this map exactly corresponds to what one would get by naively
moving the branes past each other, keeping their position in the $x_7$
direction. Thus it was not possible to confirm this particular
conjecture of \cite{uran}\footnote{See however previous footnote.}.

In conclusion one can say that we have given evidence that Toric
geometry offers a powerful tool for studying the relation between
brane configurations and branes at singularities. It would be
interesting to investigate more complicated singularitites, for
example of the type $x^{k}y^{k}=z^{k'}w^{k'}$, in an analogous way.

\vskip 1em
\noindent{\large\bf Acknowledgements}
\vskip 1em
\noindent
The author would like to thank Ulf Lindstr\"{o}m and Yuri Shirman for
useful and stimulating discussions.

\appendix
\section{Appendix}
Following the procedure outlined in the text we arrive at the
following toric data for the Higgs branch of the moduli space of the
model given in figure \ref{fig:3NS3NSp}. 
\be
T =  \left(\begin{array}{cccccccccccccccc}
2 & 2 & -1 & -1 & 1 & 1 & 1 & 1 & 1 & 1 & 0 & 0 & 0 & 0 & 0 & 0 \\
1 & 0 & 1 & 0 & 1 & 1 & 1 & 0 & 0 & 0 & 1 & 1 & 1 & 0 & 0 & 0 \\
-2 & -1 & 1 & 2 & -1 & -1 & -1 & 0 & 0 & 0 & 0 & 0 & 0 & 1 & 1 & 1
\end{array}
 \right).
\ee
It corresponds to the toric data for a space of the type
$xy=z^{3}w^{3}$ as expected. The charge matrix for the symplectic
quotient (including the FI-parameters) is given by
\be
\label{Qfor3NS}
{\tiny
Q =  \left(\begin{array}{ccccccccccccccccc}
-1 & 0 & 0 & 0 & 1 & 1 & 0 & 0 & 0 & 0 & -1 & 0 & 0 & 0 & 0 & 0 & 0\\
0 & -1 & 0 & 0 & 0 & 0 & 0 & 1 & 1 & 0 & 0 & 0 & 0 & -1 & 0 & 0 & 0\\
0 & 0 & -1 & 0 & -1 & 0 & 0 & 0 & 0 & 0 & 1 & 1 & 0 & 0 & 0 & 0 & 0\\
0 & 0 & 0 & -1 & 0 & 0 & 0 & -1 & 0 & 0 & 0 & 0 & 0 & 1 & 1 & 0 & 0\\
1 & -1 & 0 & 0 & -1 & 0 & 0 & 1 & 0 & 0 & 0 & 0 & 0 & 0 & 0 & 0 & \z_2\\
0 & 0 & 0 & 0 & -1 & 1 & 0 & 0 & 0 & 0 & 0 & 0 & 0 & 0 & 0 & 0 & \z_2
+\z_3 \\
0 & 0 & 0 & 0 & 1 & 0 & -1 & 0 & 0 & 0 & 0 & 0 & 0 & 0 & 0 & 0 & \z_6
+ \z_1 \\
0 & 0 & 0 & 0 & 0 & 0 & 0 & -1 & 1 & 0 & 0 & 0 & 0 & 0 & 0 & 0 & \z_3
+ \z_4 \\
0 & 0 & 0 & 0 & 0 & 0 & 0 & 1 & 0 & -1 & 0 & 0 & 0 & 0 & 0 & 0 & \z_1
+ \z_2\\
0 & 0 & 0 & 0 & 0 & 0 & 0 & 0 & 0 & 0 & -1 & 1 & 0 & 0 & 0 & 0 & \z_4
+ \z_5 \\
0 & 0 & 0 & 0 & 0 & 0 & 0 & 0 & 0 & 0 & 0 & -1 & 1 & 0 & 0 & 0 & \z_2
+ \z_3\\
0 & 0 & 0 & 0 & 0 & 0 & 0 & 0 & 0 & 0 & 0 & 0 & 0 & -1 & 1 & 0 & \z_5 + \z_6\\
0 & 0 & 0 & 0 & 0 & 0 & 0 & 0 & 0 & 0 & 0 & 0 & 0 & 0 & -1 & 1 & \z_3
+ \z_4
\end{array}
 \right)}
\ee

\bigskip
The toric data for the Higgs branch of the moduli space of the model
given in figure \ref{fig:dual} is the same as for the model given in
figure \ref{fig:NSconf}. The D-term equations are also the
same. However, the charge matrix implementing the symplectic quotient 
is different and is given by
\be
Q = \left(\begin{array}{ccccccccc}
0 & 0 & 0 & 0 & -1 & 1 & 0 & 0 & \z_1\\
0 & 0 & 0 & 0 & 0 & 0 & 1 & -1 & \z_3\\
0 & 1 & 0 & 1 & -2 & 0 & 0 & 0 & \z_1\\
1 & -1 & 0 & 0 & 1 & 0 & -1 & 0 & \z_2\\
-1 & 0 & -1 & 0 & 0 & 0 & 2 & 0 & \z_3
\end{array}\right).
\ee

\newcommand{\NPB}[1]{{\sl Nucl. Phys.} {\bf B#1}}
\newcommand{\PLB}[1]{{\sl Phys. Lett.} {\bf B#1}}
\newcommand{\PRL}[1]{{\sl Phys. Rev. Lett.} {\bf #1}}
\newcommand{\PRD}[1]{{\sl Phys. Rev.} {\bf D#1}}
\newcommand{\JHEP}[1]{{\sl JHEP} {\bf #1}}

\end{document}